\documentclass{nature}

\usepackage{amsmath,amssymb,amsfonts}
\usepackage{cite}
\usepackage{CJK}
\usepackage{bm}
\usepackage{tikz}
\usepackage{pgfplots}
\usetikzlibrary{matrix,fit}
\usepackage{graphicx} 
\usepackage{braket}
\usepackage{subfig}
\usepackage{caption}
\usepackage{mathrsfs}
\usepackage{float}
\usepackage[utf8]{inputenc}
\usepackage{pgflibraryarrows}
\usepackage{lipsum} 
\usepackage{pgflibrarysnakes}
\usepackage{authblk}
\usepackage{lineno}

\usepackage[labelsep=none]{caption}
\usepackage[labelfont=bf]{caption}
\usepackage[absolute,overlay]{textpos}

\title{Dimensional hierarchy of higher-order topology in three-dimensional sonic crystals}

\author{Xiujuan Zhang $^{1,2}$\thanks{These authors contributed equally to this work.} , Bi-Ye Xie$^{1,2 *}$, Hong-Fei Wang$^{1,2 *}$, Xiangyuan Xu$^{1,3}$, Yuan Tian$^{1,2}$, Jian-Hua Jiang$^{4}$\thanks{Corresponding authors: jianhuajiang@suda.edu.cn and luminghui@nju.edu.cn.} , Ming-Hui Lu$^{1,2,5,6 \dagger}$ \& Yan-Feng Chen$^{1,2,6}$}
\usepackage{graphicx}
\makeatletter
\let\saved@includegraphics\includegraphics
\AtBeginDocument{\let\includegraphics\saved@includegraphics}
\renewenvironment*{figure}{\@float{figure}}{\end@float}
\makeatother

\begin{document}
%\linenumbers
\maketitle

\begin{affiliations}
 \item National Laboratory of Solid State Microstructures, Nanjing University, Nanjing 210093, China
 \item Department of Materials Science and Engineering, Nanjing University, Nanjing 210093, China
 \item Key Laboratory of Noise and Vibration Research, Institute of Acoustics, Chinese Academy of Sciences, Beijing 100190, China.
 \item School of Physical Science and Technology, and Collaborative Innovation Center of Suzhou Nano Science and Technology, Soochow University, 1 Shizi Street, Suzhou 215006, China
 \item Jiangsu Key Laboratory of Artificial Functional Materials, Nanjing 210093, China
 \item Collaborative Innovation Center of Advanced Microstructures, Nanjing University, Nanjing 210093, China.
 \thanks
\end{affiliations}
\newpage
\begin{abstract}
Topological phases of matter have been extensively studied for their intriguing bulk and edge properties. Recently, higher-order topological insulators with boundary states that are two or more dimensions lower than the bulk states, have been proposed and investigated as novel states of matter. Previous implementations of higher-order topological insulators were based on two-dimensional (2D) systems in which 1D gapped edge states and 0D localized corner states were observed. Here we theoretically design and experimentally realize a 3D higher-order topological insulator in a sonic crystal with a large topological band gap. We observe the coexistence of third-, second- and first-order topological boundary states with codimension three, two and one, respectively, indicating a dimensional hierarchy of higher-order topological phenomena in 3D crystals. Our acoustic metamaterial goes beyond the descriptions of tight-binding model and possesses a band structure which automatically breaks the chiral symmetry, leading to the separation of bulk, surface, hinge and corner states. Our study opens a new route toward higher-order topological phenomena in three-dimensions and paves the way for topological wave trapping and manipulation in a hierarchy of dimensions in a single system.
\end{abstract}

Topological insulators (TIs) with topologically protected boundary states, going beyond the classification of the states of matter
by spontaneous symmetry breaking, have stimulated tremendous explorations in electronic~\cite{TI1,TI2}, photonic~\cite{PTI1, PTI2, PTI3, PTI4, PTI5, PTI6, PTI7, PTI8, PTI9} and phononic~\cite{PNI1, PNI2, PNI3, PNI4, PNI5, PNI6} materials. Higher-order topological insulators (HOTIs) as a new paradigm of topological states of matter exhibit unconventional bulk-boundary correspondence in which lower-dimensional boundary states can emerge~\cite{HOTI1, HOTI2, HOTI3, HOTI4, HOTI5, HOTI6, HOTI7, HOTI8, HOTI9, HOTI10, HOTI11, HOTI12, HOTI13, HOTI14, HOTI15, HOTI16}. For a $m$-dimensional ($m$D) TI, one can define the codimension of a $n$D boundary state as $l=m-n$. Then a $l$th-order TI is defined as a TI with $l$-codimensional boundary states~\cite{CO1, CO2}. Previously, the experimental realizaitons of HOTIs were focused on 2D materials by utilizing quadrupole topological insulators~\cite{HOTI1, HOTI2, HOTI3, HOTI4, HOTI15} and 2D TIs with quantized Wannier centers~\cite{HOTI10, HOTI12, HOTI13, HOTI14}.  Theoretical predictions of 3D HOTIs have been proposed in tight-binding models where only nearest-hopping between SC sites is considered~\cite{HOTI8}. However, the experimental realization of 3D HOTIs~\cite{bismuth}, especially those with coexistence of 2D surface states, 1D hinge states and 0D corner states (i.e., HOTIs with $3$-codimensional boundary states) remains challenging.

Sonic crystals (SCs) with band structures that can be designed at free will have provided an ideal platform to investigate diverse topological states such as the quantum Hall states~\cite{PNI2,PNI3,PNI6}, quantum spin Hall states~\cite{PNI1,PNI5}, crystalline topological insulating states~\cite{PNI7,HOTI12} and states with Weyl points and Fermi arcs~\cite{PNI8,PNI9}. Moreover, the unique boundary states and bulk topology in SCs can provide novel mechanisms for achieving acoustic cloaking, disorder-immune wave guiding and negative refraction~\cite{PNI7,PNI9}. With versatile techniques for acoustic wave excitation and measurements, the search for novel topological states and phenomena in SCs has attracted lots of attention in recent years and has been further facilitated by the development of 3D printing technology for the fabrication of SCs~\cite{PNI7}.

In this article, we design and fabricate a 3D SC with a large bulk band gap, with accessing toward the first-, second- and third-order topological insulating phases. As a consequence, topological surface, hinge and corner states emerge at the boundaries with different dimensions, indicating a dimensional hierarchy of higher-order topology. These topological phases are characterized by non-trivial bulk polarizations and quantizations of Wannier centers (3D Zak phases)~\cite{HOTI8, WN}. The underlying physics mimics the topological crystalline insulators~\cite{CRY, HOTIC} where the mirror symmetries restrict the positions of the Wannier centers to the maximal Wyckoff positions.

Our 3D SCs possess a simple cubic lattice geometry (lattice constant $a=2$~cm) consisting of air cavities connected by open channels, as depicted by the insets of Fig.~1a-b (the solid areas denote the hard walls while the empty spaces denote the air cavities and the air channels). Facilitated by the 3D printing technology, the photosensitive resin (serving as ``hard walls" for acoustic waves) is used to fabricate the SCs. We design two kinds of SCs which have the same bulk band structure but different topological properties. In the trivial SC (inset of Fig.~1a), an air cavity in the center is connected by six channels to the neighboring cells, while the nontrivial (topological) SC has air cavities distributed at eight corners of the unit cell (inset of Fig.~1b). These air cavities have one-eighth volume of the air cavity in the trivial SC. Such SC configurations essentially mimic the Su-Schrieffer-Heeger physics~\cite{SSH1} in three-dimensions, and thus naturally introduce the 3D Zak phase~\cite{SSH2} $(\theta_x,\theta_y,\theta_z)$ to be defined below. The band structures of the two SCs are shown in Fig.~1a-b where a full band gap emerges with a large band-gap-to-mid-gap ratio of 41\% in both cases. These two phases have different topological properties that are characterized by the parities of the eigen-states at different high-symmetric points in the first Brillouin zone (BZ) (see the insets of Fig. 1a-b). It is observed that the the parities of the eigen-states at the $\Gamma$ point are the same for the two lattices, while they exhibit opposite parity properties at the other high-symmetric points, i.e., the X, M and R points. Further illustrations on the eigen-values of the mirror symmetry operators (denoted by three values of either $1$ or $-1$) provide more evidence to the parity properties of the two phases and therefore their topological properties, i.e., the SC in Fig. 1a is topologically trivial and that in Fig. 1b is topologically nontrivial (see Supplementary Information for more details). We point out again that the band structures in our implementation have linear dispersions when the frequency approaches  zero, which are qualitatively different from the tight-binding band structures (see Supplementary Information for details).

The topological properties of our SCs are further characterized by the bulk polarizations and the quantization of the Wannier centers, similar to the recent realizations of the second-order TIs in 2D systems~\cite{HOTI8, HOTI13, HOTI14}. The 3D bulk polarization is defined as follows
\begin{equation}
p_i=-\frac{1}{(2\pi)^3}\int_{BZ} d^3\bm{k}\mathrm{Tr} [\hat{{\cal A}}_i] , \quad i=x,y,z
\end{equation}
where $(\hat{{\cal A}}_i)_{mn}(\textbf{k})=\mathrm{i}\bra{u_m(\textbf{k})}\partial_{k_i}\ket{u_n(\textbf{k})}$, with $m$, $n$ running over all bands below the acoustic band gap. $\ket{u_m(\textbf{k})}$ is the periodic part of the Bloch wavefunction for the acoustic pressure ($m$ is the band index and ${\textbf{k}}$ is the wavevector). The Wannier center (WC) is located at $(p_x, p_y, p_z)$. The 3D bulk polarization is simply related to the 3D Zak phase~\cite{SSH2} via $\theta_i=2\pi p_i$ for $i=x,y,z$. Due to the three mirror symmetries with respect to the three orthogonal axises, the WC is pinned at the maximal Wyckoff positions of the unit cell. When the WC is at the center of the unit cell, the SC is adiabatically connected to an atomic insulator which is the topologically trivial case. In contrast, the WCs of the topological crystalline insulators are away from the center of the unit cell and therefore they are also denoted as obstruct atomic insulators. There are three possible cases: i) when the WC locates at the center of the surfaces of the cubic unit cell, the insulator is a first-order TI with surface states; ii) when the WC locates at the center of the hinges of the unit cell, it is a second-order TI with both hinge states and surface states; iii) when the WC locates at the corners of the unit cell, it is a third-order TI exhibiting concurrent corner, hinge and surface states (see Supplementary Information for more details).

The bulk polarizations are related to the Wannier bands. For our SCs which have the same structure along all three directions (i.e., our SCs have mirror symmetries along all three directions), we consider only the $z$ component of the bulk polarization $p_z=\frac{1}{(2\pi)^2}\int_{S} dk_x dk_y \nu_z(k_x, k_y)$, where $\nu_z(k_x, k_y)$ represents the Wannier band~\cite{HOTI16, WN} and $S$ is the projection area of BZ on the $k_x$-$k_y$ plane. Due to the spatial symmetries, it is straightforward to deduce $p_x=p_y=p_z$. We numerically calculate the Wannier band and find that for the trivial SC, $\nu_z=0$ for all $(k_x, k_y)$ as shown in Fig.~1c, leading to $\bm{p}=(p_x, p_y, p_z)=(0, 0, 0)$. This  implies that the WC is located at the center of the unit cell and the SC is in the trivial insulating phase. For the topological SC, however, the Wannier band takes non-zero and quantized values of $\nu_z(k_x, k_y)=0.5$ as show in Fig.~1d, hence yielding a nontrivial bulk polarization of $\bm{p}=(\frac{1}{2},\frac{1}{2}, \frac{1}{2})$. This indicates that the WC is located at the corner of the unit-cell and our SC is a HOTI of the case iii). According to the above analyses, our topological SC will exhibit a dimensional hierarchy of the first-, second- and third-order topological phases. Figure 1\textbf{e} illustrates such occurrences when different types of boundaries are created. In the following, we will demonstrate these higher-order topological boundary states.    

 To investigate the first-order topological boundary states, i.e., the surface states, we numerically calculate the projected band structure (in the $k_x$-$k_y$ plane) for a ribbon-like supercell in Fig.~2a (the supercell is sketched as the inset). In-gap states (the orange curves) separated from the bulk bands are found, showing the emergence of the surface states (see Supplementary Information for the pressure filed distributions of the eigen surface states). We then fabricate a ``surface sample" (as shown in Fig.~2b), where the topological SC with $10\times10\times7$ cells is jointed to the trivial SC with the same size, to visualize the surface states. We perform three types of pump-probe measurement for detecting the bulk and surface states (see Methods for the detailed experimental set-up; see the inset of Fig.~2c for the source and detector locations). The measured transmission spectra for the bulk and surface probes are presented in Fig.~2c.It is clearly visible that the surface probe captures most acoustic energy in the frequency region of 4-7.5 kHz where the surface states are excited, while the bulk probes receive less energy as the bulk state excitation is suppressed, especially in the bulk gap region (from 7 kHz to 10 kHz). Due to the finite-size effect, the transmission spectral signature of the bulk and surface band gaps is deviated from the eigen-mode simulation. The slight frequency shifts might also originate from small fabrication errors ($\pm 0.1$ mm) and the environmental set-ups (see Methods). 

A measurement on the pressure field maps for the excited surface state is presented using the slice plot in Fig.~2d. The measurement is conducted using the same set-up as that in Fig.2c, but we fix the frequency and scan the pressure field amplitudes with a spatial resolution of $2$~mm along the $z$-direction and $2$~cm along the $x$- and $y$-direction. A frequency of 5.5~kHz is particularly chosen as it exhibits an obvious excitation peak of the surface probe, as well as the explicit suppression of the bulk probes. We observe a rapid decay of the acoustic energy away from the interface between the topological and trivial SCs, while propagation of the acoustic wave along the interface, indicating wave localization around the interface. Although the surface and bulk states share almost the same frequency range in our system (see Fig.~2a), the surface and bulk states are distinguishable in our acoustic pump-probe measurements (more evidence can be found in the Supplementary Information).        

We now study the second-order topological effects with both topological hinge and surface states. Different from the surface sample where there is only one boundary plane, in the ``hinge sample", we open boundaries along both $x$- and $y$-directions. The projected band structure of a hinge supercell on the $z$-direction is calculated and provided in Fig.~3a. A sketch of the hinge supercell is inserted, composed of a block of topological SC with $10\times10\times1$ cells surrounded by the walls of the trivial SC with the thickness of 4 cells, creating four interfaces and four hinges at their intersections. It is observed that in addition to the four degenerate surface states (the orange curves), four degenerate hinge states (the dark blue curves) are found, which are one dimension lower than the surface states, just as predicted by our theory. Notice that the numbers of the surface and hinge states are consistent with the numbers of the interfaces and hinges created. This is a numerical testimony of the bulk-surface-hinge correspondence in higher-order topology. In the Supplementary Information, we conduct a detailed study on the surface and hinge states for a particular wave vector $k_z=0$ and confirm their degeneracy. 

To visualize the bulk, surface and hinge states, we fabricate a ``hinge sample", which has the same size as the hinge supercell in the $x$-$y$ plane but a finite thickness of 10 cells along the $z$-direction (see Fig.~3b for a photograph of the sample). We then conduct the pump-probe measurements with four different configurations (see the inset of Fig.~3c). The source is placed near one of the four hinges to explicitly excite the hinge states. There are one detection for the bulk probe at the center of the sample, two detections near the interfaces for the surface probes and another detection near the hinge for the hinge probe. The measured transmission spectra are presented in Fig.~3c, where the inset illustrates the source and probe locations. Several features characterize the higher-order topology in this sample. First, we observe that the frequency range of 5.3-7.1~kHz is dominated by the hinge response, where the bulk and surface responses are considerably weaker. We further measure the acoustic pressure profile of the hinge state excited at 5.7~kHz by the same source in Fig.~3c and show the results using slice plot in Fig.~3d. The scanning step is $2$~mm along the $x$-direction and $2$~cm along the $y$- and $z$-directions. It is observed from the color maps that the acoustic energy is mostly concentrated along the hinge between the $x$-$z$ and $y$-$z$ interfaces, propagating along the $z$ direction. Away from the hinge, the acoustic energy decays quickly, showing wave localization on the hinge. The coexistence of the surface and hinge states is an important feature of the higher-order topology in three-dimensions. To further confirm this, we perform more measurements on the excitation of the surface states in the hinge sample by putting the source to a location closer to the interfaces in the Supplementary Information. The results show clear evidence of the existence of the surface states, indicating our topological SC indeed exhibits the second-order topological property.  

When all boundaries of the topological SC along all three directions are opened, the third-order topology will manifest itself as the concurrent emergence of the corner, hinge and suarface states. To verify this, we fabricate a ``corner sample", consisting of a block of topological SC with $8\times8\times8$ cells enclosed by the walls of the trivial SC with the thickness of 4 unit-cells (see Supplementary Information for a photograph of the corner sample). In order to be experimentally measurable (see Method for more details), we deliberately cut off two slices of the outer trivial SC to expose the inner topological SC (see Supplementary Information for how the cut-off is performed; see Fig.~4a for a photography of the sample used in the measurements). We conduct four different pump-probe measurements to detect the bulk, surface, hinge and corner responses (see the inset of Fig.~4b for the illustration of the positions of the source and the detection probes). The source is placed not far away from the corner, hinge and interfaces to explicitly excite these boundary states. The transmission spectra for the corner, hinge, surface and bulk states are shown in Fig.~4b. Several key-features of higher-order topology are observed. First, the appearance of a strong peak in the common spectral gap of the bulk, surface and hinge states indicates the emergence of the corner states. Second, there are direct evidences on the successive emergence of the corner, hinge and surface states in the transmission spectra, which is consistent with the physical picture of a third-order TI. Specifically, the corner probe finds a strong peak at the frequency of 7.9 kHz, the hinge probe finds peaks around 5.6 kHz, the surface probe finds peaks around 4.7 kHz and the bulk probe finds peaks around 3.4 kHz. This again confirms the above theoretical prediction on how our topological SC manifests its third-order topological properties (shown in Fig.~1e). Furthermore, the acoustic pressure field map for the excited corner state at 7.9 kHz is presented in Fig.~4c, where the typical behaviors of a localized 0D mode are observed, i.e., the mode energy is mostly concentrated around the corner while decays rapidly into the hinge, surface and bulk. The field maps for the hinge and surface states at their corresponding excited frequencies are shown in the Supplementary Information, which give clear evidences of the excitation of these higher dimensional boundary states.

We would like to point out that the above measured surface, hinge and corner states are separated from the bulk states, while this is not the case for a 3D SSH model~\cite{SSH1, SSH2, SSH3} (see details in the Supplementary Information). In the 3D SSH model, due to the alternative nearest-neighbor couplings, there exists a sublattice symmetry (more precisely, the chiral symmetry) that restricts the band structures of both unitcell and supercell structures to be symmetric with respect to the ``zero energy". As a consequence, the spectra of the topological surface, hinge and corner states are pinned to be around the zero energy. In particular, the corner states have zero energy and are always embedded in the spectra of the hinge and surface states. In contrast, our SCs naturally break the chiral symmetry and lead to the spectral separation among the topological surface, hinge and corner states.

In summary, we have realized a third-order acoustic TI with concurrent topological surface, hinge and corner states which are observed in both transmission and acoustic-field scanning measurements. This dimensional hierarchy of topological boundary states offer unprecedented topological wave trapping and manipulation that can be useful in acoustic metamaterials~\cite{TRAP}. Moreover, our SCs allow various manifestations of the higher-order topology in different dimensions (see Fig.~1e), which may be exploited for topological transfer of acoustic energy among multiple dimensions~\cite{HOTI12, C2}.

{\it Note added}: At the final stage of this work, we notice two preprints~\cite{3D1, 3D2} appeared, which realize 3D acoustic HOTIs by simulating the 3D tight-binding model in pyrochlore lattice with nearest-hopping, where different and complimentary higher-order topological phenomena are observed.

\noindent
\textbf{\large{Methods}}

\noindent
\textbf{Experiments}

\noindent
The present SCs consisting of air cavities connected by open channels are made of photosensitive resin (modulus 2,765 MPa, density 1.3 g$\cdot$~cm$^{-3}$), which serves as acoustically hard walls. A stereo lithography apparatus is used to fabricate the samples, including a surface sample, a hinge sample and a corner sample. The lattice constant of the SCs is $a$=2 cm. The geometric parameters of the cavity and channel sizes are illustrated in Fig.~1a. The geometric tolerance is roughly 0.1 mm. 

The experimental data on the transmission spectra in Figs. 2c, 3c and 4b are collected using the following procedure. We eject a point-like acoustic signal, which is generated from an acoustic transducer and guided into the sample through a thin channel made of acoustically hard material (i.e., the photosensitive resin). An acoustic detector (Knowles SPM0687LR5H MEMS  microphone with sizes of 4.72 ~mm$\times$3.76 mm) is used to probe the excited pressure field. Its position is controlled by an automatic stage and can move as required. Our SCs are carefully designed and optimized (the open channel has maximum width of 6 mm) such that there is enough space for the detector to get into the sample and probe the pressure field on demand. The data were collected and analysed using a DAQ card (NI PCI-6251). For the data in Fig.~2c, the detection positions for the bulk and surface modes are at the middle of the topological SC side, the middle of the trivial SC side and the middle of the interface. The source for excitation of the acoustic waves is placed at a position close to the interface and nearly equal distant from the three probes. For the data in Fig.~3c, a specific hinge is considered. Given the system has $C_4$ symmetry, the study on one of the four hinges is able to provide sufficient evidence to the excitation of the hinge states. In this measurement, the source is put at a location one cell away from the center of the considered hinge and the probes for the hinge, surface and bulk states are respectively performed at the center of the hinge, the center of the two interfaces intersecting at the hinge and the middle of the topological SC. For the data in Fig.~4b, similar as that for the hinge state measurements, only one corner is considered. The source is placed two cells away from the corner and the locations of the corner, hinge, surface and bulk probes are respectively at the corner, the center of the hinge along the $y$-direction, the center of the $y$-$z$ surface and the middle of the topological SC. For the transmission spectra measurements in the Supplementary Information, similar procedure is used, only the source and detecting positions are varied as required for different purposes. The said locations are all marked in the corresponding figures as insets.         

For the measurement of the boundary states, the acoustic pressure distributions (Figs.~2d, 3d and 4c, as well as those in the Supplementary Information), the above mentioned procedure is used, only the detector moves its position (controlled by the automatic stage) such that the excited modes can be spatially resolved and enough data can be collected, which are further post-processed to generate the color maps. It is worth pointing out that the straight open channels between the outer space and the inner sample space that can be accessed by the detector are separated by a lattice constant. This makes our field scanning perform with a restricted spatial step (the minimum is $a$=2 cm) along certain directions. For example, in Fig.~2d, the scanning step along the $x$- and $y$-directions is 2 cm while the scanning step along the $z$-direction can be tuned (we choose it as a fine step of 2 mm). 

There are some factors that might affect our experimental measurements and cause the discrepancies between the simulations and the experimental results. First is the finite size effect. In the simulations, periodic boundary conditions are implemented while in the experiments, the samples have finite size which can dependently shift the operating frequencies. Another factor might come from the fabrication error, which also affects the excitation of the boundary states. In addition, the physical properties of air might vary depending on the weather conditions, giving a third potential reason for the slight frequency shift between the experiments and simulations.

\noindent
\textbf{Simulations}

\noindent
Numerical simulations in this work are all performed using the 3D acoustic module of a commercial finite-element simulation software (COMSOL MULTIPHYSICS). The resin blocks are treated as acoustically rigid materials. The mass density and sound velocity in air are taken as 1.21 kg$\cdot$m$^{-3}$ and 343 m$\cdot$s$^{-1}$, respectively. In the eigen evaluations, all six boundaries of the unit cells are set as Floquet periodic boundaries for the data in Fig. 1a-d. The boundaries of the supercells are set as Floquet periodic boundaries along the boundary directions, with the perpendicular directions set as plane wave radiation boundaries, for the data in Figs. 2a and 3a. The eigen evaluations in the Supplementary Information also obey the similar set-up. In the simulations on the excitations, all boundaries are set as plane wave radiation boundaries.

%% Put the bibliography here, most people will use BiBTeX in
%% which case the environment below should be replaced with
%% the \bibliography{} command.

% \begin{thebibliography}{1}
% \bibitem{dummy} Articles are restricted to 50 references, Letters
% to 30.
% \bibitem{dummyb} No compound references -- only one source per
% reference.
% \end{thebibliography}

%\bibliographystyle{naturemag}
%\bibliography{nature-template}

%% Here is the endmatter stuff: Supplementary Info, etc.
%% Use \item's to separate, default label is ``Acknowledgements"

\begin{addendum}
 \item[Acknowledgements] X.J.Z., B.Y.X., H.F.W., X.Y.X., Y.T., M.-H. L. and Y.-F.C. are supported by the National Key R\&D Program of China (2017YFA0303702 and 2018YFA0306200) and the National Natural Science Foundation of China (grants nos. 11625418, 11890700 and 51732006). J.-H.J. is supported by the National Natural Science Foundation of China (grant no. 11675116), the Jiangsu distinguished professor funding and a project funded by the Priority Academic Program Development of Jiangsu Higher Education Institutions (PAPD). X.J.Z. thanks H. Ge and S. Yu for their support and assistance with the experimental measurements.
  
 \item[Author contributions]X.J.Z., B.Y.X and H.F.W conceived the idea. B.Y.X and H.F.W did the theoretical analyses. X.J.Z. performed the numerical simulations. X.J.Z., X.Y.X. and Y.T. performed experimental measurements. J.-H.J., M.-H.L. and Y.-F.C. guided the research. All authors contributed to discussions of the results and manuscript preparation. B.Y.X., X.J.Z., J.-H.J. and M.-H.L. wrote the manuscript.

 \item[Competing Interests] The authors declare that they have no
competing financial interests.
 %\item[Correspondence] Correspondence and requests for materials
%should be addressed to A.B.C.~(email: myaddress@nowhere.edu).
\end{addendum}

%%
%% TABLES
%%
%% If there are any tables, put them here.
%%
\pagebreak
\begin{figure}[htbp]
\begin{center}
\includegraphics[width=0.9\columnwidth]{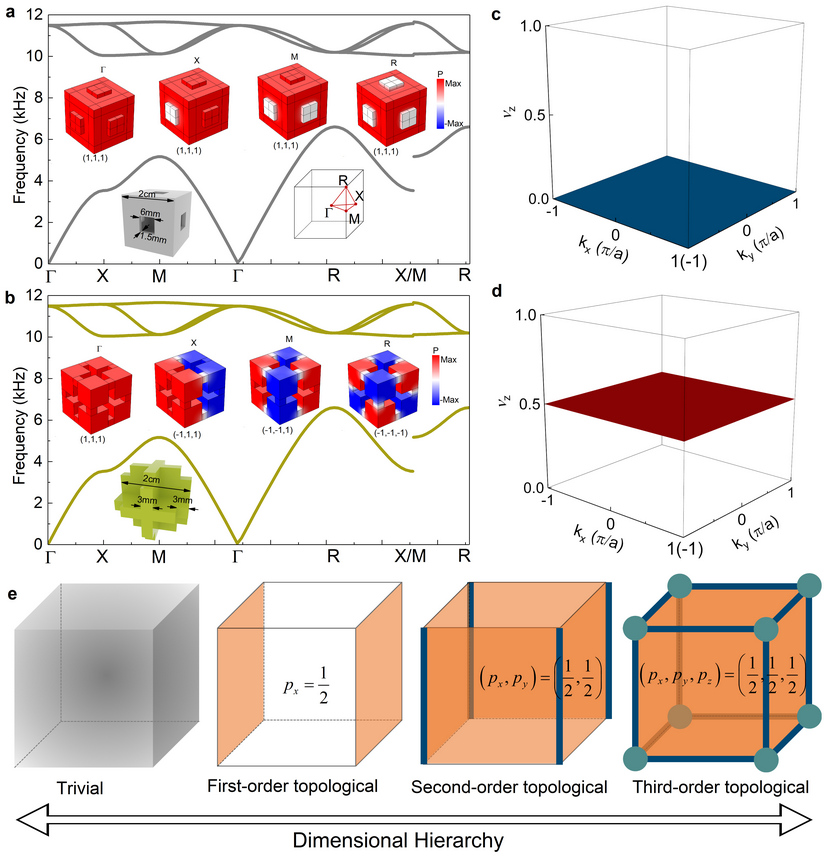}
\label{fig:1}
\end{center}
\end{figure}
\noindent
\textbf{Figure 1 $|$ Dimensional hierarchy of the higher-order topological states in 3D sonic crystals.} Simulations on the bulk band structures for the \textbf{a,} trivial and \textbf{b,} topological sonic crystals (SCs) are presented. The Brillouin zone and the unit cells of the proposed SCs with geometric parameters illustrated, are shown as insets. The SCs respect mirror symmetries along all three directions. Although the band gaps in the two SCs share the same frequency range, they carry different topological nature, i.e., one is trivial and the other is non-trivial. This is characterized by the parity properties of the eigen-states, as shown by the field maps in the insets, where the states at the $\Gamma$ point have the same parity and those at the other high-symmetric points (X, M and R points) have opposite parities (see Supplementary Information for more details). \textbf{c and d,} Wannier band calculations along $z$-direction across the $k_x$-$k_y$ plane for the trivial and topological SCs, respectively. Notice that the former carries values of 0 indicating topologically trivial bulk polarization, i.e., $p_z$=0, while the latter carries values of 0.5 indicating topologically non-trivial bulk polarization, i.e., $p_z$=0.5. Since the SCs have mirror symmetries along all three directions, it is naturally deduced $p_x=0$ and $p_y=0$ for the trivial SC and $p_x=0.5$ and $p_y=0.5$ for the topological SC. The consequences of the topological SC carrying non-trivial bulk polarization are illustrated in \textbf{e} where trivial, first-order, second-order and third-order topologies manifest themselves in hierarchical dimensions when different types of boundaries are created.

\pagebreak
\begin{figure}[htbp]
\begin{center}
\includegraphics[width=0.9\columnwidth]{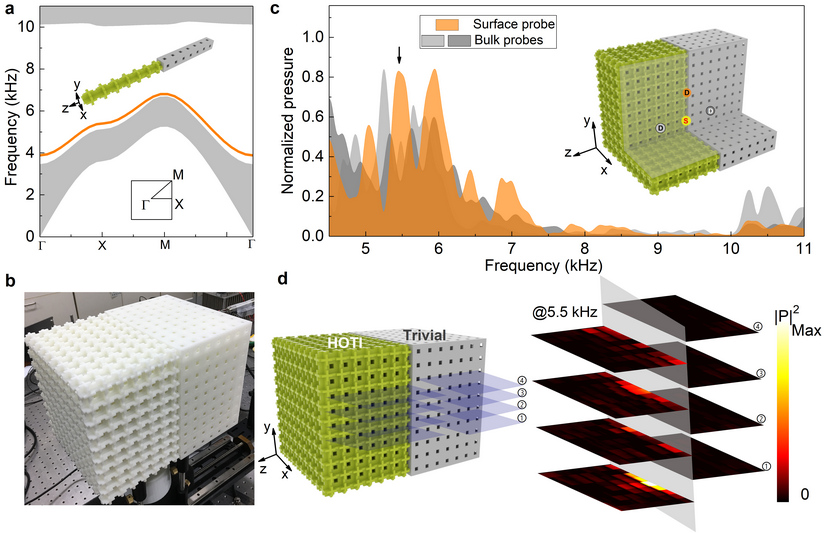}
\label{fig:2}
\end{center}
\end{figure}
\noindent
\textbf{Figure 2 $|$ First-order topological phase and the 2D surface states.} \textbf{a,} Simulated dispersions of the bulk (gray) and surface (orange) states for the supercell composed of 7 topological cells jointed with 7 trivial cells along $z$-direction (as shown by the inset). The Brillouin zone and its high-symmetric points are also depicted as inset. \textbf{b,} Photograph of the fabricated surface sample, consisting of $10\times10\times7$ topological cells jointed with the same number of the trivial cells, forming an interface on the $x$-$y$ plane. \textbf{c,} Measured transmission spectra for three types of pump-probe, including bulk from the topological SC side (light gray), bulk from the trivial SC side (dark gray) and surface (orange). Inset: sketch of the surface sample with half of its volume cut off in order to see the cross section where the locations of the source (represented by the green dot) and the probes (represented by the dots with colors corresponding to that in the main figure) are marked. \textbf{d,} Acoustic pressure profile for the surface state measured at a frequency of 5.5~kHz (whose position is also indicated in \textbf{c} by a black arrow). For the sake of clarity, a four-layered slice plot is presented, with the slicing indicated by the purple transparent surfaces. It is clearly observed that the excited mode energy is mostly concentrated on the $x$-$y$ interface and decayed along the $z$-direction, consistent with the property of a surface state.

\pagebreak
\begin{figure}[htbp]
\begin{center}
\includegraphics[width=0.9\columnwidth]{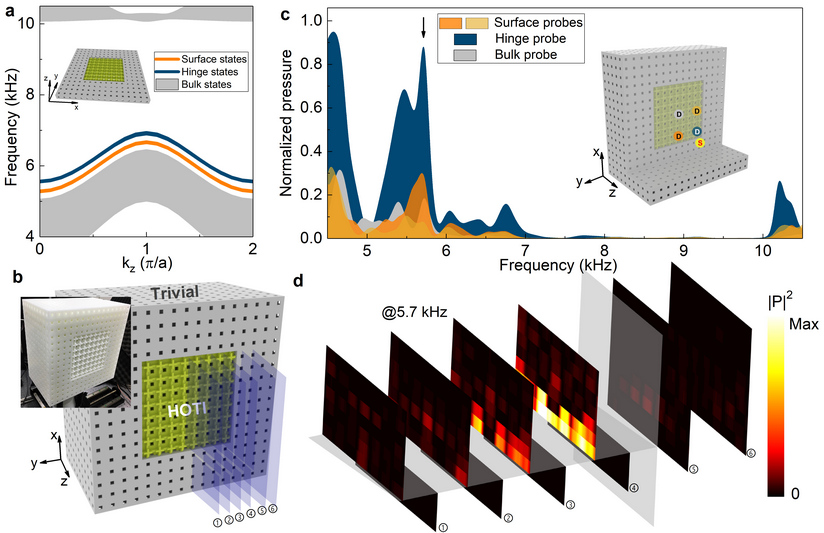}
\label{fig:3}
\end{center}
\end{figure}
\noindent
\textbf{Figure 3 $|$ Second-order topological phase and the 1D hinge states.} \textbf{a,} Simulated dispersions of the bulk (gray), surface (orange) and hinge (dark blue) states for the supercell composed of $8\times8$ topological cells surrounded by a trivial SC with thickness of 4 cells. A sketch of such a supercell is inserted. In the gap of the bulk states, there are four degenerate surface states and four degenerate hinge states obtained (see Supplementary Information for the details on the degeneracy), which are consistent with the numbers of the interfaces and hinges. \textbf{b,} Photograph of the fabricated hinge sample, consisting of $8\times8\times10$ topological cells surrounded by the trivial SC with a thickness of 4 cells, forming four interfaces and four hinges. A sketch of the hinge sample with slicing (purple transparent surfaces) is also provided to indicate the locations to conduct experimental acoustic field scanning. \textbf{c,} Measured transmission spectra for four types of pump-probe: bulk probe (gray), two surface probes (orange) and hinge probe (dark blue). Inset: the same as the inset in Fig.~2\textbf{c}, but for the hinge sample. \textbf{d,} Acoustic pressure profile for the hinge state measured at a frequency of 5.7 kHz (inicated by the black arrow in \textbf{c}). A slice plot with slicing locations indicated in \textbf{b} is presented. Different from the measured surface state in Fig.~2\textbf{d}, the excited hinge state here has mode energy mostly concentrated along the hinge between the $x$-$z$ and $y$-$z$ interfaces. When propagating away from this hinge, the acoustic wave is rapidly decayed.

\pagebreak
\begin{figure}[htbp]
\begin{center}
\includegraphics[width=0.9\columnwidth]{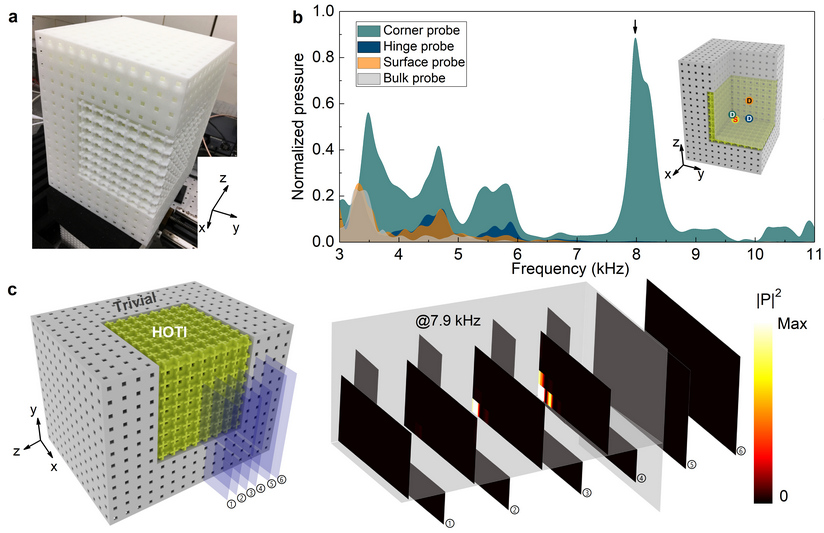}
\label{fig:4}
\end{center}
\end{figure}
\noindent
\textbf{Figure 4 $|$ Third-order topological phase and the 0D corner states.} \textbf{a,} Photograph of the fabricated corner sample, consisting of $8\times8\times8$ topological cells enclosed by the trivial SC with a thickness of 4 cells. Notice that in order to be experimentally measurable (which requires the topological SC is exposed), two areas of the corner sample are deliberately cut off (see Supplementary Information for the photograph of the complete sample and how the cut-off is performed). \textbf{b,} Measured transmission spectra for four types of pump-probe: bulk probe (gray), surface probe (orange), hinge probe (dark blue) and corner probe (dark green). Inset: the same as the insets in Fig.~2\textbf{c} and Fig.~3\textbf{c}, but for the corner sample. Notice that for the sake of clarity, the marker for the bulk probe is not shown, which locates in the middle the topological SC. \textbf{c,} Acoustic pressure profile for the corner state measured at a frequency at 7.9 kHz (indicated by the ablck arrow in \textbf{b}). A slice plot with slicing locations indicated in the sketch of the corner sample (the left panel in \textbf{c}) is presented. Typical behaviors for a localized corner state are clearly observed, where the excited mode energy is mostly concentrated at the corner and decayed rapidly when going into the hinges, surfaces and bulk.

\end{document}